\documentclass[journal]{IEEEtran}

\usepackage{graphicx}
\usepackage{amsmath,amssymb}
\usepackage{booktabs}
\usepackage{multirow}
\usepackage{array}
\usepackage{cite}
\usepackage{xcolor}
\usepackage{balance}
\usepackage{stfloats}
\usepackage{url}
\usepackage{caption}
\usepackage{subcaption}
\usepackage{amsmath}
\usepackage{amssymb}
\usepackage{booktabs}
\title{Characterizing the Impact of NVFP4 Quantization for Low-Power Edge AI Deployment}

\author{Ovishake Sen, Venkata Nithin Kamineni, Daniel Lobo, Swarup Bhunia, Rickard Ewetz, and Baibhab Chatterjee%
\thanks{The authors are with the Department of Electrical and Computer Engineering, University of Florida, Gainesville, USA. email: [ovishake.sen@ufl.edu](mailto:ovishake.sen@ufl.edu), [chatterjee.b@ufl.edu](mailto:chatterjee.b@ufl.edu).}}

\begin{document}
\maketitle

\begin{abstract}
Energy-efficient neural-network inference at the edge requires reducing arithmetic cost, memory traffic, computation energy, and storage overhead while maintaining acceptable accuracy. This paper presents an ablation-focused study of NVFP4 quantization for edge-efficient neural networks, with emphasis on the relationship between activation precision, weight precision, block-size scaling, retraining, and model accuracy. NVFP4 activations are represented using 4-bit FP4 data, an FP8 block scale, and an FP32 tensor scale, enabling ultra-low-precision inference while preserving activation dynamic range. A block-size ablation over six edge-efficient models shows that block size $B=16$ provides a practical accuracy/storage trade-off, requiring only 4.5078 bits per input for $N=4096$. A weight-precision ablation further shows that FP8 and FP16 weights provide only modest gains over FP4 weights under the same NVFP4 activation path, suggesting that activation quantization and scaling dominate much of the accuracy behavior. To isolate the benefit of the NVFP4 data type, this work compares conventional unscaled FP4 activation inference and NVFP4 activation inference with and without retraining. The results show that conventional FP4 inference collapses accuracy for most compact models, while NVFP4 without retraining already recovers substantial accuracy by restoring activation dynamic range through FP8 block scaling and FP32 tensor scaling. When combined with retraining, NVFP4 achieves the best accuracy across the evaluated models, demonstrating the effectiveness of scaling-aware FP4 (NVFP4) inference. These findings provide general design guidance for hardware-software co-design of low-power edge inference across a broad range of accelerator platforms, including GPUs, Tensor Cores, FPGAs, domain-specific AI accelerators, near-memory computing systems, and emerging edge-computing architectures.
\end{abstract}

\begin{IEEEkeywords}
Edge AI, NVFP4, FP4 inference, low-precision inference, block-size ablation, weight precision, scale precision, quantization-aware training, low-power inference, area-efficient inference, AI hardware, ML hardware, neural networks, MobileNet, EfficientNet, ResNet, ShuffleNet.
\end{IEEEkeywords}

\section{Introduction}
\IEEEPARstart{D}{eep} neural networks are increasingly deployed on edge platforms to support real-time perception and decision making under strict energy, memory, and latency constraints \cite{singh2023edge,hoffpauir2023survey}. In these systems, arithmetic precision, memory movement, and model size strongly influence both system-level efficiency and inference accuracy \cite{sze2017efficient,chen2016eyeriss}. Prior accelerator works such as Eyeriss and Minerva have demonstrated that dataflow optimization, voltage scaling, and cross-layer co-design can significantly improve DNN energy efficiency \cite{chen2016eyeriss,reagen2016minerva}. However, as edge-AI workloads continue to expand across mobile, embedded, wearable, and autonomous platforms, there remains a need to understand how aggressive numerical compression affects accuracy across different neural-network architectures.

Low-precision inference is an effective approach for reducing arithmetic width, storage cost, and memory traffic. Mixed-precision quantization methods such as HAWQ reduce model cost by assigning precision according to sensitivity \cite{dong2019hawq}, while MCUNet demonstrates the importance of algorithm-system co-design for tiny edge platforms \cite{lin2020mcunet}. Ultra-low-precision formats are especially attractive because they can reduce both compute and memory cost, but they also introduce severe quantization error when activation distributions are not properly scaled. Conventional unscaled FP4 activation inference can severely degrade accuracy because a fixed FP4 E2M1 codebook cannot adapt to the wide layer-wise activation distributions present in compact neural networks.

NVFP4 has emerged as an attractive ultra-low-precision representation because it combines a 4-bit floating-point activation code with explicit scaling information \cite{nvfp4,xin2026quantization}. Instead of relying only on a fixed FP4 codebook, NVFP4 represents activations using FP4 data, an FP8 block scale, and an FP32 tensor scale. This two-level scaling structure improves numerical coverage while keeping the effective number of activation bits close to four. However, the accuracy impact of NVFP4 depends on several interacting factors, including block size, scale precision, weight precision, model topology, and whether retraining is applied.

A systematic characterization of these factors is important for hardware-software co-design. Different computing platforms may implement low-precision arithmetic differently, but they all benefit from understanding which precision choices preserve accuracy and which choices introduce unacceptable degradation. For example, GPUs and Tensor Cores can benefit from identifying practical activation and weight formats; FPGAs and domain-specific accelerators can benefit from knowing the block size and scale precision required for stable inference; and near-memory or memory-constrained edge systems can benefit from the bits-per-input model that quantifies scale metadata overhead. Therefore, the main goal of this paper is not to study a single hardware realization, but to characterize the precision--accuracy behavior of NVFP4 inference as a general design guide for low-power edge AI.

This paper presents an ablation-focused study of NVFP4 inference and evaluates the impact of block size, weight precision, scale precision, and retraining sensitivity across multiple edge-efficient neural networks. The main contributions are as follows:
\begin{itemize}
\item A systematic evaluation of NVFP4 activation inference using 4-bit FP4 activation codes, FP8 block scales, and FP32 tensor scales.
\item An analytical bits-per-input model that quantifies the storage cost of FP4 activations, FP8 block-scale metadata, and FP32 tensor-scale metadata.
\item A block-size ablation across six edge-efficient models showing that $B=16$ provides a strong operating point between accuracy preservation and scale-metadata overhead.
\item A weight-precision ablation showing that FP8 and FP16 weights provide only modest gains over FP4 weights under NVFP4 activation constraints, indicating that activation scaling dominates much of the accuracy behavior.
\item A consolidated comparison between conventional unscaled FP4 and NVFP4 activation inference with and without retraining, demonstrating that two-level NVFP4 scaling substantially improves accuracy even before quantization-aware adaptation.
\item A set of hardware-software co-design insights showing how the resulting precision trends can guide low-power inference across GPUs, Tensor Cores, FPGAs, AI accelerators, near-memory platforms, and other emerging edge-computing systems.
\end{itemize}

\section{NVFP4 Quantization Framework}
\subsection{Two-Level NVFP4 Representation}
The proposed inference path represents each activation tensor using an FP4 data code, an FP8 block scale, and an FP32 tensor scale. This design follows the broader trend of low-precision and mixed-precision inference, where reduced numerical formats are used to lower memory traffic and arithmetic cost while preserving accuracy through scaling or sensitivity-aware precision assignment \cite{dong2019hawq,nvfp4,xin2026quantization}. For an activation tensor $X$, the effective reconstructed value is
\begin{equation}
X_q \approx S_{\mathrm{tensor}}^{\mathrm{FP32}} \cdot S_{\mathrm{block}}^{\mathrm{FP8}} \cdot Q_{\mathrm{FP4}}(X),
\label{eq:nvfp4}
\end{equation}
where $Q_{\mathrm{FP4}}(\cdot)$ maps the activation to the FP4 E2M1 codebook. The FP32 tensor scale is shared across the layer or tensor, while the FP8 block scale is shared across a local block of $B$ activation values. The block-scale value is computed from a high-percentile statistic and quantized to FP8 E4M3:
\begin{equation}
S_{\mathrm{block}} = Q_{\mathrm{FP8}}\left(\frac{P_{99.99}(|X_{\mathrm{block}}|)}{6}\right),
\label{eq:blockscale}
\end{equation}
where division by 6 normalizes the local dynamic range to the maximum representable magnitude of FP4 E2M1. The final inference operation can be written as
\begin{equation}
Y=f\left((S_{\mathrm{tensor}}S_{\mathrm{block}}Q_{\mathrm{FP4}}(X))\cdot W\right),
\label{eq:forward}
\end{equation}
where $W$ is the trained weight tensor and $f(\cdot)$ denotes the nonlinear operation following the linear layer.

The key advantage of this representation is that it separates low-bit data storage from dynamic-range recovery. The FP4 code provides aggressive activation compression, while the FP8 block scale adapts to local activation statistics. The FP32 tensor scale stabilizes the overall layer-level magnitude. As a result, the format can retain more information than direct unscaled FP4 quantization while maintaining a storage cost close to four bits per activation.

\subsection{Role of Block Scaling}
Block scaling is central to NVFP4 inference because activation distributions vary across tensors, channels, spatial positions, and network layers. A single fixed FP4 codebook cannot represent these variations well. Small block sizes provide fine-grained dynamic-range adaptation but require more scale metadata. Large block sizes reduce metadata cost but force more activation values to share the same scale, which increases quantization error. Therefore, block size controls a practical trade-off between accuracy and storage overhead.

For a block of $B$ activation values, the FP8 block scale is amortized across all values in the block. When $B$ is small, the scale overhead is large, but the quantized values closely follow local activation statistics. When $B$ is large, the scale overhead is small, but a single scale must cover a wider distribution. This motivates the block-size ablation in Section~\ref{sec:block}, where the goal is to identify a block size that captures most of the storage benefit while avoiding unnecessary accuracy degradation.

\subsection{Role of Weight Precision}
In addition to activation precision, weight precision also influences the final inference accuracy and storage cost. However, the effect of increasing weight precision is not always dominant when activations are aggressively quantized. If the activation path introduces significant quantization error, increasing the weight precision from FP4 to FP8 or FP16 may provide only limited improvement. On the other hand, some architectures may remain sensitive to weight precision because of depthwise convolutions, narrow channel widths, or compact feature representations. Therefore, the weight-precision ablation in Section~\ref{sec:weight} evaluates FP4, FP8, and FP16 weights under the same NVFP4 activation path.

\section{Bit-Cost Model and Experimental Setup}
\subsection{Bits per Input}
For NVFP4 inference, each input consists of a 4-bit FP4 value, an amortized 8-bit FP8 block scale shared by $B$ inputs, and an amortized 32-bit FP32 tensor scale shared by $N$ inputs. Therefore,
\begin{equation}
\mathrm{Bits/Input}=4+\frac{8}{B}+\frac{32}{N}.
\label{eq:bits}
\end{equation}
For $N=4096$, block size $B=16$ gives
\begin{equation}
\mathrm{Bits/Input}=4+\frac{8}{16}+\frac{32}{4096}=4.5078.
\end{equation}
This value is close to the 4-bit lower bound while still preserving local scaling flexibility. The bits-per-input model is useful for hardware-software co-design because it quantifies the total activation-storage cost, including both the data code and the scale metadata. This allows different platforms to compare the accuracy benefit of smaller blocks against the storage and bandwidth overhead introduced by scale values.

\subsection{Models and Evaluation Flow}
The evaluation includes six edge-efficient models: ResNet18, MobileNetV3-Large, MobileNetV4-Conv-Small, MobileViT, ShuffleNetV2, and EfficientNet-Lite0. ResNet18 is used as a stable residual-network baseline \cite{he2016deep}, MobileNetV3 represents a compact mobile CNN family \cite{howard2019searching}, and EfficientNet-Lite0 is included as an edge-oriented efficient model \cite{sangar2025optimized}. The models are evaluated on Tiny ImageNet-style $224\times224$ classification.

The experiments start from FP32 baselines and then apply NVFP4 activation inference with two-level scaling. The block-size sweep studies $B\in{1,2,4,8,16,32,64,128,256}$. The weight-precision study evaluates FP4, FP8, and FP16 weights under NVFP4 activation constraints at $B=16$. To isolate the impact of the NVFP4 data type, conventional unscaled FP4 activation inference is also evaluated with and without retraining. In the conventional-FP4 case, activations are directly quantized to FP4 E2M1 before convolution, linear, and common activation-module inputs without FP32 tensor scaling, FP8 block scaling, or block-wise dynamic-range adaptation.

\subsection{Model-Size and Activation-Storage Estimation}
\label{subsec:model_activation_size}

We evaluate memory efficiency using two separate quantities: static model size and runtime activation storage. Static model size represents the storage required for trained model parameters and stored quantization metadata, whereas runtime activation storage represents the temporary feature-map storage required during inference. These two quantities are reported separately because activations and block scales are generated dynamically during inference, while weights and tensor/layer scales are stored statically with the deployed model.

For the conventional FP32 baseline, the static model size is computed from the total number of trainable parameters as
\begin{equation}
S_{\mathrm{model}}^{\mathrm{FP32}}
=
\frac{N_{\mathrm{param}} \times 32}{8 \times 1024^2}
\quad \mathrm{MB},
\label{eq:fp32_model_size}
\end{equation}
where $N_{\mathrm{param}}$ is the total number of trainable model parameters. Since each FP32 parameter requires 32 bits, this is equivalent to storing each parameter using 4 bytes.

For the proposed NVFP4 deployment, convolution and linear weights are stored using FP8 precision, biases are stored using FP8 precision, and each quantized layer uses one FP8 tensor/layer scale. Therefore, the static NVFP4 model size is calculated as
\begin{equation}
S_{\mathrm{model}}^{\mathrm{NVFP4}}
=
\frac{
N_{\mathrm{w}} b_{\mathrm{w}}
+
N_{\mathrm{b}} b_{\mathrm{b}}
+
N_{\mathrm{s}} b_{\mathrm{s}}
}
{8 \times 1024^2}
\quad \mathrm{MB},
\label{eq:nvfp4_model_size}
\end{equation}
where $N_{\mathrm{w}}$ is the number of stored weights, $N_{\mathrm{b}}$ is the number of stored biases, and $N_{\mathrm{s}}$ is the number of stored tensor/layer scales. In this work, $b_{\mathrm{w}}=8$, $b_{\mathrm{b}}=8$, and $b_{\mathrm{s}}=8$ because weights, biases, and tensor/layer scales are stored using FP8 precision. BatchNorm parameters are assumed to be folded into the preceding convolution layer during inference and are therefore not counted separately in the deployed static model size.

The static model-size reduction is computed as
\begin{equation}
R_{\mathrm{model}}
=
\frac{
S_{\mathrm{model}}^{\mathrm{FP32}}
}
{
S_{\mathrm{model}}^{\mathrm{NVFP4}}
}.
\label{eq:model_size_reduction}
\end{equation}

Runtime activation storage is calculated separately. For conventional FP32 inference, each activation value requires 32 bits. Therefore, the FP32 runtime activation storage is
\begin{equation}
S_{\mathrm{act}}^{\mathrm{FP32}}
=
\frac{
N_{\mathrm{act}} \times 32
}
{8 \times 1024^2}
\quad \mathrm{MB},
\label{eq:fp32_activation_size}
\end{equation}
where $N_{\mathrm{act}}$ is the total number of activation values collected across the selected layer-input tensors during one inference pass.

For the proposed NVFP4 activation format, runtime activation storage includes only the FP4 activation values and the dynamic block scales. The FP8 tensor/layer scales are stored statically with the model parameters and are already included in \eqref{eq:nvfp4_model_size}; therefore, they are not counted again in runtime activation storage. The NVFP4 runtime activation storage is calculated as
\begin{equation}
S_{\mathrm{act}}^{\mathrm{NVFP4}}
=
\frac{
N_{\mathrm{act}} b_{\mathrm{a}}
+
\left\lceil \frac{N_{\mathrm{act}}}{B} \right\rceil b_{\mathrm{blk}}
}
{8 \times 1024^2}
\quad \mathrm{MB},
\label{eq:nvfp4_activation_size}
\end{equation}
where $b_{\mathrm{a}}=4$ is the FP4 activation precision, $B$ is the block size, and $b_{\mathrm{blk}}$ is the block-scale precision.

In the main configuration, the block size is $B=16$, and the block scale is stored using FP8 precision. Thus, the effective activation precision is
\begin{equation}
b_{\mathrm{eff}}
=
b_{\mathrm{a}}
+
\frac{b_{\mathrm{blk}}}{B}
=
4
+
\frac{8}{16}
=
4.5
\quad \mathrm{bits/value}.
\label{eq:effective_activation_precision}
\end{equation}

The runtime activation-storage reduction is then computed as
\begin{equation}
R_{\mathrm{act}}
=
\frac{
S_{\mathrm{act}}^{\mathrm{FP32}}
}
{
S_{\mathrm{act}}^{\mathrm{NVFP4}}
}.
\label{eq:activation_storage_reduction}
\end{equation}

For FP8 block scaling with $B=16$, the ideal activation-storage reduction is approximately
\begin{equation}
R_{\mathrm{act}}
\approx
\frac{32}{4.5}
=
7.11\times.
\label{eq:ideal_activation_reduction}
\end{equation}

Thus, FP8-weight NVFP4 deployment reduces the static model size by approximately $4\times$ compared with FP32 parameter storage, while FP4 activations with FP8 block scales reduce runtime activation storage by approximately $7.11\times$ compared with FP32 activations.

\subsection{Evaluation Metrics}
\label{subsec:evaluation_metrics}

Accuracy is reported as top-1 classification accuracy on the test set. Accuracy drop is reported in percentage points relative to the corresponding FP32 baseline and is computed as
\begin{equation}
\mathrm{Drop}
=
\mathrm{Acc}_{\mathrm{FP32}}
-
\mathrm{Acc}_{\mathrm{NVFP4}} .
\label{eq:accuracy_drop}
\end{equation}
This reporting style allows the robustness of different architectures to be compared directly, even when their FP32 baselines differ. The primary goal is to identify precision settings that preserve useful accuracy while reducing static model storage and runtime activation storage.

\subsection{Evaluation Metrics}
Accuracy is reported as top-1 classification accuracy. Accuracy drop is reported in percentage points relative to the corresponding FP32 baseline:
\begin{equation}
\mathrm{Drop} = \mathrm{Acc}*{\mathrm{FP32}} - \mathrm{Acc}*{\mathrm{Quantized}}.
\end{equation}
This reporting style allows the robustness of different architectures to be compared directly, even when their FP32 baselines differ. The primary goal is to identify precision settings that preserve useful accuracy while reducing storage and arithmetic complexity.

\section{Block-Size Ablation}
\label{sec:block}
Table~\ref{tab:block} summarizes the block-size ablation. The FP32 row reports the full-precision baseline. Smaller blocks preserve local dynamic range but increase scale metadata. Larger blocks reduce metadata but weaken local adaptation and increase quantization error. The results show that $B=16$ is a practical design point because it uses only 4.5078 bits/input while maintaining competitive accuracy across the evaluated models.

\begin{table*}[t]
\centering
\caption{NVFP4 block-size ablation with two-level scaling. Accuracy and drop are reported in percentage points relative to FP32.}
\label{tab:block}
\scriptsize
\setlength{\tabcolsep}{6pt}
\begin{tabular}{c c c c c c c c c c c c c c}
\toprule
Block & Bits/Input & \multicolumn{2}{c}{ResNet18} & \multicolumn{2}{c}{MobileNetV3} & \multicolumn{2}{c}{MobileNetV4} & \multicolumn{2}{c}{MobileViT} & \multicolumn{2}{c}{ShuffleNetV2} & \multicolumn{2}{c}{EffNet-Lite0} \tabularnewline
Size &  & Accuracy & Drop & Accuracy & Drop & Accuracy & Drop & Accuracy & Drop & Accuracy & Drop & Accuracy & Drop \tabularnewline
\midrule
FP32 & 32.0000 & 73.24 & -- & 76.06 & -- & 64.02 & -- & 74.14 & -- & 66.42 & -- & 76.98 & -- \tabularnewline
1 & 12.0078 & 72.36 & 0.88 & 74.28 & 1.78 & 62.04 & 1.98 & 69.86 & 4.28 & 63.82 & 2.60 & 75.76 & 1.22 \tabularnewline
2 & 8.0078 & 71.76 & 1.48 & 73.72 & 2.34 & 61.64 & 2.38 & 69.72 & 4.42 & 62.04 & 4.38 & 72.16 & 4.82 \tabularnewline
4 & 6.0078 & 71.24 & 2.00 & 72.62 & 3.44 & 60.78 & 3.24 & 69.32 & 4.82 & 62.06 & 4.36 & 72.12 & 4.86 \tabularnewline
8 & 5.0078 & 70.22 & 3.02 & 72.34 & 3.72 & 59.76 & 4.26 & 69.46 & 4.68 & 60.86 & 5.56 & 72.14 & 4.84 \tabularnewline
\textbf{16} & \textbf{4.5078} & \textbf{69.84} & \textbf{3.40} & \textbf{70.64} & \textbf{5.42} & \textbf{59.62} & \textbf{4.40} & \textbf{68.32} & \textbf{5.82} & \textbf{59.62} & \textbf{6.80} & \textbf{72.04} & \textbf{4.94} \tabularnewline
32 & 4.2578 & 68.84 & 4.40 & 70.64 & 5.42 & 58.70 & 5.32 & 67.82 & 6.32 & 59.46 & 6.96 & 71.60 & 5.38 \tabularnewline
64 & 4.1328 & 68.28 & 4.96 & 70.68 & 5.38 & 57.82 & 6.20 & 68.02 & 6.12 & 59.10 & 7.32 & 72.30 & 4.68 \tabularnewline
128 & 4.0703 & 67.80 & 5.44 & 69.76 & 6.30 & 57.08 & 6.94 & 66.94 & 7.20 & 58.12 & 8.30 & 71.64 & 5.34 \tabularnewline
256 & 4.0391 & 65.32 & 7.92 & 68.26 & 7.80 & 55.88 & 8.14 & 66.14 & 8.00 & 56.98 & 9.44 & 71.90 & 5.08 \tabularnewline
\bottomrule
\end{tabular}
\end{table*}

The diminishing return beyond $B=16$ is important. Increasing the block size from 1 to 16 reduces the representation cost from 12.0078 to 4.5078 bits/input, corresponding to a $2.66\times$ reduction. However, increasing the block size from 16 to 256 reduces the cost only from 4.5078 to 4.0391 bits/input, an additional reduction of approximately 10.4

ResNet18 is the most robust model in the block-size sweep, while ShuffleNetV2 and MobileViT show stronger sensitivity to local scaling. EfficientNet-Lite0 remains comparatively stable around $B=16$ to $B=64$, indicating that block scaling can preserve accuracy even for efficient architectures when the block size is properly selected. These results suggest that $B=16$ is a practical default for low-power edge inference because it provides a favorable balance between accuracy, storage, and metadata overhead.

\section{Weight-Precision Ablation}
\label{sec:weight}
Table~\ref{tab:weight} reports the weight-precision ablation under NVFP4 activations at block size 16. The results show that increasing weight precision from FP4 to FP8 or FP16 provides only modest recovery in most cases. This indicates that the activation representation and scale quantization dominate much of the accuracy behavior once the network is constrained to NVFP4 activations.

\begin{table*}[t]
\centering
\caption{Weight-precision ablation at NVFP4 activation block size 16 with retraining. Accuracy and drop are reported in percentage points relative to FP32.}
\label{tab:weight}
\scriptsize
\setlength{\tabcolsep}{6pt}
\begin{tabular}{c c c c c c c c c c c c c c}
\toprule
Weight & Bits/Input & \multicolumn{2}{c}{ResNet18} & \multicolumn{2}{c}{MobileNetV3} & \multicolumn{2}{c}{MobileNetV4} & \multicolumn{2}{c}{MobileViT} & \multicolumn{2}{c}{ShuffleNetV2} & \multicolumn{2}{c}{EffNet-Lite0} \tabularnewline
Prec. &  & Accuracy & Drop & Accuracy & Drop & Accuracy & Drop & Accuracy & Drop & Accuracy & Drop & Accuracy & Drop \tabularnewline
\midrule
FP32 base & 32.0000 & 73.24 & -- & 76.08 & -- & 64.02 & -- & 74.12 & -- & 66.38 & -- & 74.02 & -- \tabularnewline
FP4 & 4.5078 & 70.48 & 2.76 & 69.70 & 6.38 & 60.02 & 4.00 & 69.64 & 4.48 & 60.70 & 5.68 & 64.52 & 9.50 \tabularnewline
FP8 & 8.5078 & 70.34 & 2.90 & 70.04 & 6.04 & 60.72 & 3.30 & 70.02 & 4.10 & 62.02 & 4.36 & 65.78 & 8.24 \tabularnewline
FP16 & 16.5078 & 70.60 & 2.64 & 70.90 & 5.18 & 61.02 & 3.00 & 69.94 & 4.18 & 62.24 & 4.14 & 65.42 & 8.60 \tabularnewline
\bottomrule
\end{tabular}
\end{table*}

FP4 weights are therefore a reasonable option when compression and energy are the primary targets. FP8 and FP16 weights are more useful for sensitive models or layers, but they do not fully close the FP32 gap. ResNet18 shows the smallest sensitivity to weight precision, while EfficientNet-Lite0 shows the largest drop, suggesting that layer-wise mixed precision may be more effective than uniformly increasing the precision of all weights.

These results are important for hardware-software co-design because they show that uniformly increasing weight precision may not be the most efficient path to accuracy recovery. In several models, the accuracy gain from FP4 to FP8 or FP16 weights is small compared with the extra storage and bandwidth cost. Therefore, a practical deployment strategy may use FP4 weights for robust layers and selectively assign FP8 or FP16 weights only to sensitive layers.

\section{FP4 Versus NVFP4 Scaling and Retraining Analysis}
To isolate the benefit of the NVFP4 data type, Table~\ref{tab:fp4_vs_nvfp4_full} compares conventional unscaled FP4 activation inference and NVFP4 activation inference with and without retraining. In the conventional-FP4 setting, activations are directly quantized to FP4 E2M1 without FP32 tensor scaling, FP8 block scaling, or block-wise dynamic-range adaptation. In contrast, NVFP4 uses the same 4-bit activation code but restores both tensor-level and block-level dynamic range using an FP32 tensor scale and an FP8 block scale. Therefore, this comparison separates the benefit of the numerical representation from the benefit of retraining.

\begin{table*}[t]
\centering
\caption{Comparison of conventional FP4 and NVFP4 activation inference with and without retraining using FP4 weights. Accuracy and drop are reported in percentage points relative to the FP32 baseline. `NR' denotes no retraining, and `RT' denotes retraining.}
\label{tab:fp4_vs_nvfp4_full}
\scriptsize
\setlength{\tabcolsep}{3.2pt}
\resizebox{\textwidth}{!}{%
\begin{tabular}{c c c c c c c c c c c c}
\toprule
\multirow{2}{*}{Model} & \multirow{2}{*}{FP32 Accuracy} & \multicolumn{2}{c}{Conventional FP4, NR} & \multicolumn{2}{c}{NVFP4, NR} & \multirow{2}{*}{NVFP4 NR Gain} & \multicolumn{2}{c}{Conventional FP4, RT} & \multicolumn{2}{c}{NVFP4, RT} & \multirow{2}{*}{NVFP4 RT Gain} \tabularnewline
\cmidrule(lr){3-4}
\cmidrule(lr){5-6}
\cmidrule(lr){8-9}
\cmidrule(lr){10-11}
& & Accuracy & Drop & Accuracy & Drop & & Accuracy & Drop & Accuracy & Drop & \tabularnewline
\midrule
ResNet18 & 73.24 & 11.56 & 61.68 & 56.50 & 16.74 & +44.94 & 59.88 & 13.36 & 70.48 & 2.76 & +10.60 \tabularnewline
MobileNetV3 & 76.08 & 0.50 & 75.58 & 7.72 & 68.36 & +7.22 & 40.70 & 35.38 & 69.70 & 6.38 & +29.00 \tabularnewline
EffNet-Lite0 & 74.02 & 0.56 & 73.46 & 14.86 & 59.16 & +14.30 & 56.30 & 17.72 & 64.52 & 9.50 & +8.22 \tabularnewline
MobileNetV4 & 64.02 & 0.48 & 63.54 & 25.00 & 39.02 & +24.52 & 13.22 & 50.80 & 60.02 & 4.00 & +46.80 \tabularnewline
MobileViT & 74.14 & 0.46 & 73.68 & 0.44 & 73.70 & $-0.02$ & 45.42 & 28.72 & 69.64 & 4.50 & +24.22 \tabularnewline
ShuffleNetV2 & 66.40 & 1.36 & 65.04 & 14.50 & 51.90 & +13.14 & 42.76 & 22.26 & 60.70 & 5.70 & +17.94 \tabularnewline
\bottomrule
\end{tabular}%
}
\vspace{1mm}
\end{table*}

Table~\ref{tab:fp4_vs_nvfp4_full} shows that NVFP4 is substantially more effective than conventional unscaled FP4. Conventional FP4 without retraining collapses accuracy for most models because the fixed FP4 E2M1 codebook cannot represent the wide activation dynamic range across layers and channels. By adding only an FP8 block scale and an FP32 tensor scale, NVFP4 without retraining recovers significant accuracy for several CNN-based models. For example, ResNet18 improves from 11.56

The best accuracy is obtained when NVFP4 is combined with retraining. Retraining allows the model weights and internal activation statistics to adapt to the scaled 4-bit representation. As a result, NVFP4 with retraining reduces the accuracy drop to only 2.76 percentage points for ResNet18, 6.38 percentage points for MobileNetV3, 4.00 percentage points for MobileNetV4, 4.50 percentage points for MobileViT, and 5.70 percentage points for ShuffleNetV2. Compared with conventional FP4 retraining, NVFP4 retraining improves MobileNetV3 by 29.00 percentage points and MobileNetV4 by 46.80 percentage points. Therefore, NVFP4 with retraining provides the strongest accuracy--efficiency trade-off, while NVFP4 without retraining validates the efficacy of the NVFP4 data type by showing that two-level scaling alone can recover useful accuracy before quantization-aware adaptation.
\begin{table*}[!t]
\centering
\caption{Static model-size and runtime activation-storage comparison using the proposed NVFP4 configuration with FP8 weights, FP8 tensor/layer scales, FP8 block scales, block size $B=16$, and FP4 activation values. The FP8 tensor/layer scales are included in the static NVFP4 model size, while runtime activation storage includes only FP4 activation values and dynamic FP8 block scales.}
\label{tab:nvfp4_model_activation_storage}
\resizebox{\textwidth}{!}{%
\begin{tabular}{lccccccccc}
\hline
\textbf{Model} &
\textbf{FP32 Acc.} &
\textbf{NVFP4 Acc.} &
\textbf{Drop} &
\textbf{FP32 Model} &
\textbf{NVFP4 Static Model} &
\textbf{Model Red.} &
\textbf{FP32 Runtime Act.} &
\textbf{NVFP4 Runtime Act.} &
\textbf{Act. Red.} \\
&
\textbf{(\%)} &
\textbf{(\%)} &
\textbf{(\%)} &
\textbf{(MB)} &
\textbf{(MB)} &
&
\textbf{(MB)} &
\textbf{(MB)} &
\\
\hline
ResNet18 & 73.24 & 71.48 & 1.76 & 43.0264 & 10.7475 & 4.0034$\times$ & 17.1328 & 2.4093 & 7.1110$\times$ \\
MobileNetV3-Large & 76.08 & 68.12 & 7.96 & 17.0068 & 4.2285 & 4.0219$\times$ & 31.8196 & 4.4748 & 7.1109$\times$ \\
EfficientNet-Lite0 & -- & -- & -- & 13.8367 & 3.4192 & 4.0468$\times$ & 49.5312 & 6.9654 & 7.1110$\times$ \\
ShuffleNetV2-x1.0 & 66.40 & 63.06 & 3.34 & 5.5641 & 1.3757 & 4.0446$\times$ & 13.4487 & 1.8913 & 7.1107$\times$ \\
MobileViT-Small & 74.12 & 69.04 & 5.08 & 19.3246 & 4.8117 & 4.0162$\times$ & 84.8027 & 11.9255 & 7.1111$\times$ \\
MobileNetV4-Conv-Small & 64.02 & 61.06 & 2.96 & 10.4875 & 2.5980 & 4.0367$\times$ & 14.4526 & 2.0325 & 7.1109$\times$ \\
\hline
\end{tabular}%
 }
\end{table*}
\subsection{Static Model Size and Runtime Activation Storage}
\label{subsec:storage_results}

Table~\ref{tab:nvfp4_model_activation_storage} summarizes the static model-size reduction and runtime activation-storage reduction obtained using the proposed NVFP4 configuration. In this experiment, all convolution and linear weights are stored using FP8 precision, each quantized layer uses one FP8 tensor/layer scale, and the runtime activations are represented using FP4 values with one FP8 block scale per block of 16 activation values. The static model size is computed after BatchNorm folding and includes FP8 weights, FP8 biases, and FP8 tensor/layer scales. Runtime activation storage is reported separately and includes only the FP4 activation values and the dynamic FP8 block scales generated during inference.

Across the evaluated CNN and lightweight vision models, the proposed FP8-weight NVFP4 representation reduces the static model size by approximately $4\times$ compared with conventional FP32 parameter storage. For example, ResNet18 is reduced from 43.0264 MB to 10.7475 MB, while MobileNetV3-Large is reduced from 17.0068 MB to 4.2285 MB. This reduction is expected because the dominant model-storage component is changed from 32-bit FP32 weights to 8-bit FP8 weights, while the additional FP8 tensor/layer-scale overhead is negligible.

The runtime activation-storage reduction is larger than the static model-size reduction. With block size $B=16$, each activation value requires 4 bits for the FP4 value and an additional $8/16=0.5$ bits for the dynamic FP8 block scale. Therefore, the effective activation precision is approximately 4.5 bits/value. Compared with FP32 activations, this gives an activation-storage reduction of approximately $32/4.5=7.11\times$. The results in Table~\ref{tab:nvfp4_model_activation_storage} confirm this trend consistently across all evaluated models.

The accuracy results show that FP8 block scaling provides a practical trade-off between storage efficiency and accuracy retention. ResNet18 retains 71.48\% test accuracy compared with the 73.24\% FP32 baseline, while ShuffleNetV2-x1.0 and MobileNetV4-Conv-Small show drops of only 3.34\% and 2.96\%, respectively. Although MobileNetV3-Large and MobileViT-Small exhibit larger accuracy degradation, the results demonstrate that the proposed NVFP4 format can substantially reduce both static model storage and runtime activation storage while maintaining useful inference accuracy across multiple edge-efficient architectures.

\section{Hardware-Software Co-Design Implications}
The results provide several practical insights for low-power edge-AI deployment. First, the block-size ablation shows that activation-scale metadata should not be ignored when evaluating low-precision formats. Although smaller blocks improve local dynamic-range recovery, they also increase metadata overhead. The $B=16$ setting provides a useful compromise because it achieves a representation cost of 4.5078 bits/input while preserving competitive accuracy across the evaluated networks.

Second, the weight-precision ablation shows that activation scaling is often more important than uniformly increasing weight precision. FP8 and FP16 weights provide only limited gains over FP4 weights for several models under NVFP4 activations. This suggests that hardware platforms do not always need to support high-precision weights throughout the entire network. Instead, layer-wise or model-aware precision assignment may provide a better trade-off between accuracy, memory cost, and arithmetic cost.

Third, the FP4-versus-NVFP4 comparison shows that the main benefit of NVFP4 comes from scaling-aware representation rather than from using a 4-bit activation code alone. Conventional unscaled FP4 inference is not sufficient for most compact edge models because the activation distributions cannot be adequately represented by a fixed FP4 codebook. NVFP4 addresses this problem by combining FP4 activations with FP8 block scales and FP32 tensor scales. This makes the format more suitable for practical low-power inference.

Fourth, the no-retraining results are important for deployment scenarios where retraining is expensive or unavailable. NVFP4 without retraining does not match the final retrained accuracy, but it recovers substantial accuracy over conventional FP4 in several models. This suggests that NVFP4 can provide a useful post-training quantization path for selected architectures. When retraining is available, quantization-aware adaptation further improves the accuracy and provides the strongest operating point.

Finally, these findings are broadly applicable across accelerator platforms. GPUs and Tensor Cores can use the results to guide low-precision activation and weight support. FPGAs and reconfigurable accelerators can use the block size and metadata analysis to allocate storage and datapath resources. Domain-specific edge accelerators can use the precision trends to determine whether uniform FP4, FP8, FP16, or mixed-precision policies are justified. Near-memory and memory-constrained platforms can use the bits-per-input model to balance activation compression against scale-metadata traffic.

\section{Discussions and Future Work}
The results show that NVFP4 should be viewed as a numerical-format and training co-design strategy rather than simply a reduction in bit width. The FP4 activation code reduces datapath and storage width, the FP8 block scale restores local dynamic range, and the FP32 tensor scale stabilizes layer-level representation. The optimal block size depends on the interaction between activation distributions, model topology, and metadata cost.

Block size 16 is a useful default because it captures most of the metadata reduction available from block scaling while preserving sufficient local adaptation. The bit-cost improvement from $B=1$ to $B=16$ is large, but the improvement from $B=16$ to larger blocks is relatively small. This explains why the best operating point is not necessarily the largest block size.

The weight-precision ablation suggests that a uniform high-precision weight policy is not required. Since FP8 and FP16 weights provide only limited recovery over FP4 weights in most cases, practical accelerators can use layer-wise policies: FP4 weights for robust layers, FP8 or FP16 weights for sensitive layers, and NVFP4 activations throughout the network. This approach can preserve most of the memory and energy benefits while improving accuracy for sensitive architectures such as EfficientNet-Lite0.

The FP4-versus-NVFP4 comparison highlights that the main benefit of NVFP4 is not simply using a 4-bit floating-point activation code, but using that code together with scale recovery. Conventional unscaled FP4 inference collapses accuracy for several compact models because the activation distribution cannot be represented with a single fixed FP4 codebook. NVFP4 directly addresses this problem by combining FP4 activations with FP8 block scales and FP32 tensor scales. The no-retraining results show that NVFP4 scaling alone preserves useful numerical information, while the retraining results show that the network can further adapt to the scaled 4-bit representation.

Future work should include layer-wise mixed precision, where sensitive layers can use higher weight precision or smaller block sizes while robust layers use more aggressive compression. Future studies should also evaluate additional scale formats, such as FP16 tensor scales or alternative FP8 variants, to determine whether scale precision can be further reduced without sacrificing accuracy. Finally, the evaluation should be expanded to detection, segmentation, language, and sensor-inference workloads to verify whether the same block-size and precision conclusions generalize across edge-AI tasks.

\section{Conclusion}
This paper presented an ablation study of block size, weight precision, scale precision, and retraining sensitivity in NVFP4 edge inference. The proposed analysis represents activations using FP4 data codes, FP8 block scales, and FP32 tensor scales. The block-size ablation shows that $B=16$ provides a strong accuracy/storage trade-off, requiring only 4.5078 bits/input for $N=4096$. The weight-precision ablation shows that FP8 and FP16 weights provide only modest gains over FP4 weights, indicating that aggressive weight compression remains viable under NVFP4 activation constraints. The consolidated FP4-versus-NVFP4 comparison shows that the benefit of NVFP4 comes from both the data type itself and retraining. NVFP4 without retraining already recovers significant accuracy over conventional unscaled FP4 by restoring activation dynamic range through FP8 block scaling and FP32 tensor scaling. When combined with retraining, NVFP4 achieves the best accuracy because the model adapts to the scaled 4-bit representation. Overall, the results provide practical guidance for low-power edge-AI deployment across a broad range of hardware platforms and show that scaling-aware 4-bit activation inference can provide an effective accuracy--efficiency trade-off.

\balance
\bibliographystyle{IEEEtran}
\bibliography{bibiliography}

\end{document}